\documentstyle[12pt]{article}

\input{tcilatex}
\begin{document}

\author{A. de Souza Dutra\thanks{%
E-mail: dutra@feg.unesp.br}, V. G. C. S. dos Santos and A. M. Stuchi \\
UNESP-Campus de Guaratinguet\'a-DFQ\\
Av. Dr. Ariberto Pereira Cunha, 333\\
C.P. 205\\
12500-000 Guaratinguet\'a SP Brasil}
\title{{\LARGE Method of approximation for potentials in impenetrable boxes:
Harmonic Oscillator and Coulomb potentials.}}
\maketitle

\begin{abstract}
In this work we develop an approach to obtain analytical expressions for
potentials in an impenetrable box. It is illustrated through the particular
cases of the harmonic oscillator and the Coulomb potential. In this kind of
system the energy expression respect the correct quantum limits, which is a
very important quality. The similarity of this kind of problem with the
quasi exactly solvable potentials is explored in order to accomplish our
goals.
\end{abstract}

Quantum systems under non trivial boundary conditions, corresponding to
penetrable and nonpenetrable walls, do simulate the effect of atoms or
molecules in the neighbor of a central particle. The dependence of the
eigenvalues on the box size, allows one to define the effect of the pressure
over the system. Some examples would include that of the proton-deuteron
transformation as a source of energy in dense stars \cite{auluck}, the
determination of the escape rate from galactic and globular clusters \cite
{chandrasekar}, and the understanding of the spectral line shift under
pressure \cite{degroot}. One good list of phenomena associated to this kind
of physical system is presented in \cite{froman}. Recently, the fabrication
of the quantum dots in the semiconductor physics did brings a renewed
interested on this kind of problem \cite{reed}. Another example of
application of this type of system, would be that of the Casimir effect and
even that of Rydberg atoms between metallic plates \cite{marrocco}.

These important systems are treated by imposing the vanishing of the wave
function in some finite point, once that the confined particle must be found
in a finite region of the space. On the other hand, a numerical treatment of
this problem has the natural drawback of becoming difficult any qualitative
analysis, and even estimation, due to the need of calculating the energy for
each new value of the potential parameter \cite{aguilera}, \cite{goldman}.
Another possibility of studying the system is to apply to the
Rayleigh-Schroedinger perturbation theory \cite{fernandez}, \cite{fernandez2}%
, variational methods \cite{brownstein}, \cite{marin}, or some semiclassical
approximation \cite{dutt}. Recently it was applied the strong-coupling
expansion approach to this kind of problem \cite{kleinert}, and considering
the case of the pressure exerted by a stack of membranes upon enclosing
walls \cite{kleinert1}, \cite{netz}, it has been treated particularly the
case of a single membrane \cite{kleinert2}, \cite{kleinert3}.

Here we show that the imposition of the vanishing of the wave function at
the boundaries imply that the potential can be treated in a similar fashion
to that of the so called quasi exactly solvable potentials \cite{shiffman}-%
\cite{ushveridze}, which are quantum potentials which have part of the
energy spectrum exactly solvable, provided that some relations between the
potential parameters hold. In fact, we use this feature in order to deal
with the problem in the same way that it was done in the case of an
anharmonic oscillator \cite{castro}, \cite{castro2}. As an application of
the method, we present an analytically approximated expression for the
quantum problem of one particle under the action of a harmonic oscillator
bounded to a finite region of the space. Furthermore, our analytical
expression does have all the correct physical limits as, for instance, that
for large values of the principal quantum number. On the other hand, in
contrast with other approaches, our result is valid for arbitrary values of
the box size.

First of all, we illustrate the method by treating the harmonic oscillator
bounded in a ``one-dimensional box'' of length $a$. Some similar problems
were treated previously in the literature \cite{aguilera}, \cite{marin}, but
in general only solutions valid for a specific value of the potential
parameter are taken and, every time one needs the energy spectra for another
value of that parameter, the numerical calculation must be repeated. This
implies that one hardly can visualize a qualitative behavior of the system,
besides there is the obvious unpleasant need of repeated numerical
calculations each time one needs to change the value of the potential
parameter. By the other hand, methods like the perturbation theory are
intrinsically limited due to the need of restricting the potential parameter
or the length of the box.

\section{Harmonic Oscillator in a one dimensional box}

We start by treating the harmonic oscillator in a box, similarly to the
treatment given to the quasi-exactly solvable potentials \cite{shiffman}-%
\cite{ushveridze}, which did lead us to obtain analytically approximated
solutions for the eigenvalues of anharmonic oscillators under usual boundary
conditions \cite{castro}.

Remembering that the general solution for the harmonic oscillator is given
by something like

\begin{equation}
\Psi _{n}\left( x\right) =P_{n}\left( x\right) \exp \left( -\alpha
x^{2}\right) ,
\end{equation}

\noindent where P$_{n}$(x), in principle, is a polynomial of infinite degree
(the parabolic cylindrical functions), which however is taken finite
(Hermite polynomials) in the case of boundary conditions at the infinity, in
order to get a convergent function. Here however, the exponential does not
guarantees that the wavefunction vanishes at the walls of the interval, so
that this condition determines the energy spectrum. Notwithstanding, when we
do insist to using finite polynomials, which are nothing but the known
Hermite ones, whose energy is well defined and given by E$_{n}=\left[ \hbar
\,\omega \left( n+\frac{1}{2}\right) \right] $, we get an equation for the
oscillator frequency which, once solved, give us those frequencies which
generate polynomials whose zeroes are at the walls of the box. At this point
it is important to stress the similarity of this problem with that of the so
called quasi-exactly solvable potentials, in the sense that in this case
only for some specific frequencies the eigenfunction will be elementary or,
in other words, it is represented by a finite polynomial, the Hermite ones,
times an usual exponentially decaying factor. Once one gets those
frequencies, one must to substitute them on the above expression for the
energy, so obtaining the corresponding exact energy for the harmonic
oscillator in a box.

In order to obtain the frequency and the energy for an oscillator in a box
whose walls are at -a/2 and a/2, by applying the idea which was outlined
above, we use the condition

\begin{equation}
H_{n}\left( \pm \frac{a}{2}\right) =0,  \label{hermite}
\end{equation}

\noindent so that $H_{0}(x)=1$, does not give us any solution, the
polynomial $H_{2}\left( \pm \frac{a}{2}\right) $ generates one solution, $%
H_{4}\left( \pm \frac{a}{2}\right) $ two solutions, so for and so on. To be
more precise, let us exemplify with case of $H_{2}\left( \pm \frac{a}{2}%
\right) $. In this case the equation which implies that the wavefunction
vanishes at the boundary is given by 
\begin{equation}
H_{2}\left( \pm \frac{a}{2}\right) =\,a_{0}\left( -2+4\,\xi \right) \,=\,0\,,
\end{equation}
where $\xi \,=\,\sqrt{\frac{m\,\omega }{\hbar }}$. Remembering that in this
case $E_{2}=\frac{5}{2}\hbar \,\omega $, one gets finally: 
\begin{equation}
\omega =\,\frac{2\,\hbar }{m\,a^{2}},\,E\,=5\,\frac{\hbar ^{2}}{m\,a^{2}}.
\end{equation}
$\,$As can be easily verified, the wavefunction of the harmonic oscillator
for the above frequency and energy vanishes at the boundaries, not
presenting any nodes between the walls. So it corresponds to the ground
state of the oscillator in a box with impenetrable walls. Now, if one takes
the following even case of $H_{4}\left( \pm \frac{a}{2}\right) $, it is
straightforward to show that there are now two possible solutions of \ref
{hermite}, respectively corresponding to the ground state and the second
excited one, so for and so on$.$ Analogously a similar structure appears in
the odd cases. The first solution of each polynomial always corresponds to
the first excited state, the second to the third excited state, etc...

However, before to do such a calculation, we take profit of a scaling
symmetry of this kind of problem in order to simplify it. In doing so, we
show that this problem has only one independent variable for the energy
dependence. By choosing $x=\,\frac{a}{2}\,\sqrt{\frac{\hbar }{m\,\omega _{a}}%
}y$, and redefining the frequency and the energy conveniently, one gets 
\begin{equation}
-\frac{1}{2}\frac{d^{2}\psi }{dy^{2}}\,+\,\frac{1}{2}\,\,\omega
_{a}^{2}\,y^{2}\,\psi \,=\,E_{a}\,\psi ,
\end{equation}

\noindent where the new variables in terms of the old ones look like 
\begin{equation}
\omega _{a}\,=\,\frac{m}{\hbar }\frac{a^{2}}{4}\,\omega ,\,\,E_{a}\,=\,\frac{%
ma^{2}}{4\,\hbar ^{2}}\,E,
\end{equation}

\noindent and now we need to find the zeros of $\psi $ such that $\psi
\left( \pm \,1\right) =\,0$. The resulting solution will be valid for
arbitrary values of the box length and its frequency. In this way, by
solving a sufficiently great number of polynomials, we can plot the energy
as a function of the frequency for each one of the energy levels, and then
try to get a functional relation between these several levels. So obtaining
an approximate analytical expression for the dependence on the frequency,
the length of the box and the quantum number.

At this point, in order to help our quest for a suitable analytical
expression for the energy, we remember that in the limit of zero frequency
one recalls the free particle in a box. On the other hand, when the length
of the box goes to infinity, one should recover the well known harmonic
oscillator spectra. Finally for great values of the principal quantum
number, once again the energy becomes close to that of a free particle in a
box. As a consequence of these physical constraints, we write the expression
for the energy as 
\begin{equation}
{\cal E\,}\,=\,\left( \frac{2\,m\,a^{2}}{\pi ^{2}\hbar ^{2}}\right)
E\,=\,n^{2}\,+\,k\,\omega _{a}\,+\,g\left( \omega _{a}\right) ,
\end{equation}

\noindent where 
\begin{equation}
k\,=\,\frac{8}{\pi ^{2}}\left( n\,-\,\frac{1}{2}\right) ,\,n=1,2,3,...
\end{equation}

\noindent and the function $g\left( \omega _{a}\right) $ should have the
following appropriate limits 
\begin{equation}
g\left( \omega _{a}\,=\,0\right) \,=\,0,\,g\left( \omega _{a}\,\rightarrow
\,\infty \right) \,=\,0.
\end{equation}

\noindent Furthermore, one should also have this function going to zero when 
$n$ increases or at most growing more slowly than $n^{2}$. Using this
hypothesis as a guide we try the following form for $g_{n}\left( \omega
_{a}\right) $: 
\begin{equation}
g_{n}\left( \omega _{a}\right) \,=\,c_{0}\left( n\right) \,\omega
_{a}\,e^{-\sum_{j=1}^{J}\,c_{j}\left( n\right) \,\omega _{a}^{j}}.
\end{equation}

After performing the necessary calculations we verify that indeed, this
function has the expected behavior supposed above. Besides, by obtaining the
roots of Hermite polynomials up to two hundred degree, we did find an
approximated expression for the coefficients $c_{j}\left( n\right) $, taking 
$J=\,3$, whose expressions are given below: 
\begin{equation}
c_{0}\left( n\right) \,=\,\frac{1}{\left( 0.405231\,+\,0.810579\,n\right) },
\end{equation}
\begin{equation}
c_{1}\left( n\right) \,=\,\frac{0.0104832}{n}\,-\,\frac{0.00588616}{n^{2}}%
\,-\,\frac{0.00187449}{n^{3}},
\end{equation}
\begin{equation}
c_{2}\left( n\right) \,=\,10^{-6}\left[ -1.24+1.35\,\cos h\left(
0.1762\,\,n\,-\,0.12\right) \right] ^{-1},\,n>1,
\end{equation}
\begin{equation}
c_{3}\left( n\right) \,=10^{-8}\left[ -2.5+2.6\,\cos h\left(
0.086\,\,n\,-\,0.278\right) \right] ^{-1},\,n>3,
\end{equation}

\noindent Note that for the last two coefficients, the first elements were
separated in order to getting better fittings. In these cases one have

\[
c_{2}\left( 1\right) \,=\,3.70973\,10^{-6}\,, 
\]
\begin{equation}
c_{3}\left( 1\right) \,=-1.54146\,10^{-6}\,,\,c_{3}\left( 2\right)
\,=-8.78283\,10^{-7}\,,\,c_{3}\left( 3\right) \,=\,4.56951\,10^{-8}\,.
\end{equation}

Note that, in fact, the coefficients really approaches to zero for larger
values of $n$. These calculations were performed for $n\leq \,20$, so that
for higher values of $n$, one must extrapolate it, but it should be expected
good results due to the behavior of $g_{n}\left( \omega _{a}\right) $ when $%
n $ becomes greater and greater.

The comparison of the energy coming from the above analytical approximation
with pure numerical values, in the case of the range of frequencies
considered, shows that the error was always less than $0.07\%$ along the
range of the parameters verified which due to technique reasons, depend of
the energy level studied and grows for higher quantum numbers, below we
present a Table with the corresponding ranges and their respective maximum
percentual errors for the first twenty levels. 
\[
\begin{array}{llllll}
n & \omega _{a_{\max }} & \delta \%_{\max } & n & \omega _{a_{\max }} & 
\delta \%_{\max } \\ 
1 & 2.0 & 9.0\times 10^{-5} & 11 & 60.5 & 0.018 \\ 
2 & 6.0 & 0.072 & 12 & 67.3 & 0.018 \\ 
3 & 10.9 & 9.1\times 10^{-3} & 13 & 74.1 & 0.017 \\ 
4 & 16.3 & 0.02 & 14 & 81 & 0.017 \\ 
5 & 22.1 & 0.016 & 15 & 87.9 & 0.016 \\ 
6 & 28.1 & 9.3\times 10^{-5} & 16 & 94.9 & 0.016 \\ 
7 & 34.4 & 0.012 & 17 & 101.9 & 0.015 \\ 
8 & 40.8 & 0.017 & 18 & 109 & 0.013 \\ 
9 & 47.2 & 0.019 & 19 & 116.1 & 0.016 \\ 
10 & 53.8 & 0.019 & 20 & 123.2 & 8.9\times 10^{-3}
\end{array}
\]

It is interesting to stress that the type of energy expression obtained
here, can be used to study the appearing of quantum chaos in the cases of
anharmonic systems \cite{gutzwiller}, \cite{hu}. This happens because one
can use the energy expressions coming from this approach in order to get the
number of levels for a given range of energy and the corresponding related
parameters, used to discuss the appearing of the quantum chaos.

\section{\protect\bigskip \protect\smallskip Coulomb Potential in a sphere}

In this section we work with the more mathematically involved case of the
Coulomb potential in a sphere of radius $a.$ Notwithstanding, the basic
program to be followed is the same of the previous section. Once again the
general solution for the radial wavefunction of this problem is of the kind

\begin{equation}
R(x)=R_{n}^{*}(x)\exp \left( \frac{-\mu \alpha r}{\hbar ^{2}n}\right)
\label{1}
\end{equation}

\noindent where V(r) = -$\alpha /r$, and $n$ is the principal quantum
number. This polynomial will be of infinity degree, unless that we impose
that the wavefunction vanishes at the infinity, in which case the polynomial
degree is finite, one gets the generalized Laguerre polynomials. However, as
in the previous section, despite we are treating a case with finite boundary
conditions for which it has in general an infinity number of terms, we will
insist in using a finite polynomial in order to generate simple exact
solutions, remembering that the price to be paid is that consequently, the
energy is established and given by $E_{n}=-\mu \alpha ^{2}/\left( 2\,\hbar
\,n\right) $. As a consequence the constant $\alpha $ will be chosen in a
manner to guarantee the vanishing of the wavefunction at the sphere surface.
After that we substitute the so obtained constant in the above expression
for the energy, getting the corresponding energy for the system.

Starting from the radial Schroedinger equation, once that the angular part
of it is unchanged due to the fact that the boundary condition preserves the
spherical symmetry \cite{schiff}:

\begin{equation}
\frac{-\hbar ^{2}}{2\mu }\frac{1}{r}\frac{d^{2}}{dr^{2}}\left[ r\left(
R\left( r\right) \right) \right] +\left[ \frac{l\left( l+1\right) }{2\mu
r^{2}}\hbar ^{2}+V\left( r\right) \right] R\left( r\right) =E\,R\left(
r\right)  \label{2}
\end{equation}

\noindent with $V(r)=-\frac{e^{2}}{r}$. Defining $R(r)=\frac{u\left(
r\right) }{r}$, we have\ 

\begin{equation}
\left[ \frac{-\hbar ^{2}}{2\mu }\frac{d^{2}}{dr^{2}}+\frac{l\left(
l+1\right) }{2\mu r^{2}}\hbar ^{2}+V\left( r\right) \right] u\left( r\right)
=E\,u\left( r\right) .  \label{3}
\end{equation}

\noindent Redefining the radial variable as $\rho =\frac{\mu e^{2}}{\hbar
^{2}}r$ and then substituting it in the previous equation one gets

\begin{equation}
\left[ C\left[ -\frac{d^2}{d\rho ^2}+\frac{l\left( l+1\right) }{\rho ^2}%
-\frac 2\rho \right] -E\right] v\left( \rho \right) =0  \label{4}
\end{equation}

\noindent where $C=\frac{\mu e^{4}}{2\hbar ^{2}}$. By choosing the ansatz u$%
\left( \rho \right) =v\left( \rho \right) \exp \left( -ar\right) $ one
obtains

\begin{equation}
\left[ C\left[ -\frac{d^{2}}{d\rho ^{2}}+2a\frac{d}{d\rho }-a^{2}+\frac{%
l\left( l+1\right) }{\rho ^{2}}-\frac{2}{\rho }\right] -E\right] v\left(
\rho \right) =0.  \label{5}
\end{equation}

\noindent By choosing $v\left( \rho \right) =\rho
^{l+1}\sum_{0}^{n}b_{n}\rho ^{n}$ and performing the identification $%
E=-Ca^{2},$ we get the following recurrence relation\smallskip \ 

\begin{equation}
b_{n+1}=\frac{2\left( 1-a\left( n+l+1\right) \right) }{l(l+1)-(n+l+2)(n+l+1)}%
b_{n}.  \label{6}
\end{equation}

\noindent Taking into account the boundary conditions at the infinity, we
can truncate the series at $b_{N}$ ($N$, integer), getting $b_{N+1}=0$,
so\smallskip \ obtaining

\[
a=\frac{1}{N+l+1}, 
\]

\noindent and as $E=-Ca^{2}$, one gets the usual expression for the
eigenenergies in such boundary conditions\smallskip \ 

\begin{equation}
E_{N}=-\frac{C}{(N+l+1)^{2}}=-\frac{\mu e^{2}}{2\hbar ^{2}}\frac{1}{n^{2}}.
\label{8}
\end{equation}

We can still use the recurrence relation in order to obtain an expression
for the generic coefficient $b_{n}$ in terms of $b_{0}$. In doing so we get\ 

\begin{equation}
\frac{b_{n}}{b_{0}}=\prod_{j=l}^{n}\left\{ \frac{2\left[ a\left( j+l\right)
-1\right] }{j(j+2l+1)}\right\} .  \label{9}
\end{equation}

\smallskip \noindent As in the case of boundary condition at a finite
radius, the series does not truncate and so we must to work with the
infinite polynomial generated from the above coefficients. This polynomial
will be analogous to the Parabolic Cylinder function which appeared in the
case of the harmonic oscillator, as treated in the last section.\noindent

On the other hand the equation for $R(r)$, can be written in terms of the
Laguerre polynomial\smallskip \ 

\begin{equation}
R(r)=r^{l}\exp \left( -kr\right) L_{n-l-1}^{2l+1}\left( 2kr\right) ,
\label{10}
\end{equation}

\smallskip \noindent where $k=\sqrt{\frac{2\mu E}{\hbar ^{2}}}$. Imposing
that $R(r=r_{0})=0$, we get\smallskip \ 

\begin{equation}
L_{n-l-1}^{2l+1}\left( \frac{2\mu e^{2}r_{0}}{\hbar ^{2}n}\right)
=L_{n-l-1}^{2l+1}\left( \frac{g}{n}\right) \,=\,0,  \label{11}
\end{equation}

\noindent with $g=\frac{2\mu \,e^{2}r_{0}}{\hbar ^{2}}$. In absolute analogy
with the case of the harmonic oscillator, in this case we must to substitute
the values of $g$ coming from the above equation in that of the energy of
the Coulomb potential,\smallskip \ 

\begin{equation}
E_{n,l}=-\,\frac{\mu e^{4}}{2\hbar ^{2}n^{2}}\,=-\frac{g_{n,l}^{2} 
\rlap{\protect\rule[1.1ex]{.325em}{.1ex}}h%
^{2}}{8n^{2}\mu \,r_{0}^{2}}.  \label{12}
\end{equation}
\smallskip \ 

Now, by using the recurrence relation and imposing that the polynomial be
finite, one obtains as expected the usual generalized Laguerre polynomials.
In this case, the first solution coming from the equation \ref{11} for each
polynomial will give us the ground state energy, each second solution
generates the first excited state, so for so on. Consequently we used the
Mathematica software in order to perform this calculation in a systematic
way. This was done for the first nine energy levels, using one hundred
degree polynomials. For each energy level, we plotted the exact energies
coming from the solutions described above as a function of the parameter $g$%
, and fitted it with a polynomial of the third degree given by

\begin{equation}
E_{n,l}=\sum_{m=0}^{3}C_{m}^{\left( n,l\right) }g^{m}  \label{14}
\end{equation}

\noindent where the coefficients $C_{m}^{\left( n,l\right) }$ are presented
in the Table 1.

\smallskip\ 

\[
\begin{tabular}{lllll}
n,l\TEXTsymbol{\backslash}m & 0 & 1 & 2 & 3 \\ 
1,0 & -1.00421 & 0.502846 & 0.0152158 & 0.00487229 \\ 
2,0 & -4.50591 & 0.918442 & 0.0402514 & 0.00162719 \\ 
2,1 & -2.05892 & 0.384483 & 0.00167988 & 0.000448289 \\ 
3,0 & -9.69984 & 0.938899 & -0.0182804 & 0.000282908
\end{tabular}
\]

The energy expression coming from the use of the above parameters is in good
accordance with the exact numerical data and, as happened in the harmonic
oscillator case, does have a increasing range of validity with the
increasing of the principal quantum number, as can be seen in the Table
below. In these ranges, the error is quite small, about $10^{-2}\%$. 
\[
\begin{tabular}{llll}
n,l & g$_{\max }$ & g$_{\min }$ & $\delta \,\%\,_{\max }$ \\ 
1,0 & 2.0 & 1.8 & 4.7$\times $10$^{-5}$ \\ 
2,0 & 7.1 & 6.15 & 4.3$\times 10^{-3}$ \\ 
2,1 & 6.0 & 5.0 & 6.2$\times 10^{-4}$ \\ 
3,0 & 15.5 & 12.9 & 9.4$\times 10^{-3}$%
\end{tabular}
\]

\noindent By analyzing the data, one can verify that the range was limited
by the maximum polynomial used (in this case we used polynomials of degree
100). So, in order to increase the precision and the range of validity, one
should use higher degrees for the Laguerre polynomials.

One interesting feature observed was that, in this case the energy is no
more degenerate in the angular momentum quantum number, in contrast with the
usual boundary condition at the infinite. So indicating the broke of some
symmetry of the system. As the spherical symmetry was preserved by the
boundary condition used, we must to look for another broken symmetry to be
responsible for this behavior. In fact, the symmetry related to the
conservation of the Runge-Lenz vector is that one. This can be argued by
remembering that classically is associated to fact that the eccentricity of
the elliptical orbits does not alter their energy. However in a finite
sphere some of those will not be allowed. From the point of view of a
geometric transformation, one can use the fact that the commutator of the
Runge-Lenz vector with the radial coordinate, will generate a change in this
coordinate such as\ 

\begin{equation}
r^{\prime }=r+\{r,{\bf \alpha .E}\},\,\,\,\,\,\,\left\{ r,\alpha .{\bf E}%
\right\} =\frac{1}{km}\frac{\left( {\bf rxL}\right) }{r}.{\bf \alpha ,}
\label{16}
\end{equation}

\noindent \noindent where ${\bf E}$ and ${\bf \alpha }$ are respectively the
Runge-Lenz vector and the infinitesimal vectorial parameter of this symmetry
transformation. From above one can conclude that when one is close to the
frontier, the transformation could led it to beyond it, and this is not
allowed by the boundary condition. Consequently this symmetry is now broken
at the quantum level. It is interesting to observe that technically the
Runge-Lenz operator and the Hamiltonian of the system is still commuting
but, now due to the boundary conditions, the eigenfunctions of the
Hamiltonian of this complete set of commuting operators is not shared with
those of the Runge-Lenz operator. It is interesting to note that this is a
generic property of this kind of system.

On the other hand it is not difficult to extend this approach in order to
take care of other situations like, for instance, the case of finite walls,
symmetric and antisymmetric boundary conditions, which can represent other
kind of physical systems. Finally it should be interesting to study the
applicability of this approach to systems with time-dependent boundary
conditions \cite{razavy}-\cite{marchewka}.

\smallskip\ 

\noindent {\bf Acknowledgments:} The authors are grateful to FAPESP\ and
CNPq for partial financial support.


\bigskip

\begin{thebibliography}{99}
\bibitem{auluck}  F. C.Auluck, Proc. Natl. Inst. Sci. India. {\bf 7 }(1941)
133.

\bibitem{chandrasekar}  S. Chandrasekar, Astroph. J. {\bf 97} (1943) 263.

\bibitem{degroot}  S. R. De Groot e C. A. Ten Seldam, J. Math. Physics
(Utrecht) {\bf 4} (1937) 981.

\bibitem{froman}  P. O. Fr\"{o}man, S. Yngve e N. Fr\"{o}man, J.Math. Phys. 
{\bf 28} (1987) 1813.

\bibitem{reed}  M. A. Reed, Sci. Am.{\bf \ 268} (1993) 118.

\bibitem{marrocco}  M. Marroco, M. Weidinger, R. J. Sang e H. Walther, Phys.
Rev. Lett. {\bf 81} (1998) 5784.

\bibitem{aguilera}  V. C. Aguilera-Navarro, H. Iwamoto, E. Ley Koo and A. H.
Zimerman, Il Nuovo Cimento {\bf 62} (1981) 91.

\bibitem{goldman}  S. Goldman and C. Joslin, J. Phys. Cem. {\bf 96} (1992)
6021.

\bibitem{fernandez}  F. M. Fernandez and E. A. Castro, Phys. Rev. A {\bf 46}
(1992) 7288.

\bibitem{fernandez2}  F. M. Fernandez, Phys. Rev. A {\bf 48} (1993) 189.

\bibitem{brownstein}  K. R. Brownstein, Phys. Rev. Lett. {\bf 71} (1993)
1427.

\bibitem{marin}  J. L. Marin and S. A. Cruz, Am J. Phys. {\bf 56} (1988)
1134.

\bibitem{dutt}  R. Dutt, A. Mukherjee e Y. P. Varshni, Phys. Rev. A {\bf 52}
(1995) 1750.

\bibitem{kleinert}  H. Kleinert, A. Chervyakov and B. Hamprecht, Phys. Lett.
A {\bf 260} (1999) 182.

\bibitem{kleinert1}  W. Janke, H. Kleinert and H. Meinhardt, Phys. Lett. B 
{\bf 217} (1989) 525.

\bibitem{netz}  R. R. Netz and R. Lipowski, Europhys. Lett. {\bf 29} (1995)
345.

\bibitem{kleinert2}  H. Kleinert, Phys. Lett. A {\bf 257} (1999) 269.

\bibitem{kleinert3}  M. Bachmann, H. Kleinert and A. Pelster, Phys. Lett. A 
{\bf 261} (1999) 127.

\bibitem{shiffman}  M. A. Shiffman, Int. J. Mod. Phys. A {\bf 4} (1989) 2897
and {\bf 4} (1989) 3305.

\bibitem{boschi}  A. de Souza Dutra and H. Boschi Filho, Phys. Rev. A {\bf 44%
} (1991) 4721.

\bibitem{ushveridze}  A. G. Ushveridze, Quasi-Exactly Solvable Models in
Quantum Mechanics, Institute of Physics, Bristol, 1994, and references
therein.

\bibitem{castro}  A. de Souza Dutra, A. S. de Castro and H. Boschi Filho,
Phys. Rev. A {\bf 51} (1995) 3480.

\bibitem{castro2}  A. S. de Castro and A. de Souza Dutra, Phys. Lett. A {\bf %
269} (2000) 281.

\bibitem{gutzwiller}  M. C. Gutzwiller, Chaos in Classical and Quantum
Mechanics, Springer-Verlag, New York, 1990.

\bibitem{hu}  B. Hu, B. Li, J. Li and Y. Gu, Phys. Rev. Lett. {\bf 82,}
(1999) 4224.

\bibitem{razavy}  R. Razavy, Phys. Rev. A {\bf 44} (1991) 2384.

\bibitem{brown}  L. S. Brown, Phys. Rev. Lett. {\bf 66}\ (1991) 527.

\bibitem{marchewka}  A. Marchewka and Z. Schuss, Phys. Rev. A {\bf 61}
(2000) 2107.

\bibitem{schiff}  L. I. Schiff, Quantum Mechanics. Editora McGraw-Hill Book
Company e Kogakusha Company.
\end{thebibliography}
\end{document}